\documentclass[aps,prb,twocolumn,floatfix,letterpaper]{revtex4}
\usepackage{amsmath}
\usepackage{epsfig}
\usepackage{graphicx,color,floatflt,mathrsfs,float}
\usepackage{mathptmx}
\begin{document}

\title{Hartree-Fock calculations of a finite inhomogeneous quantum wire}
\author{Jiang Qian$^1$ and Bertrand I. Halperin$^1$}

\affiliation{$^1$ Lyman Laboratory of Physics, Harvard University, 
Cambridge, MA 02138, USA}
\begin{abstract}
We use the Hartree-Fock method to study an interacting one-dimensional 
electron system on a finite wire, partially depleted at the center by a smooth 
potential barrier. A uniform one-Tesla Zeeman field is applied throughout the 
system.  We find that with the increase in the potential barrier, the low 
density electrons under it go from a non-magnetic state to an antiferromagnetic 
state, and then to a state with a well-localized spin-aligned region isolated 
by two antiferromagnetic regions from the high density leads.  At this final 
stage, in response to a continuously increasing barrier potential, the system 
undergoes a series of abrupt density changes, corresponding to the successive 
expulsion of a single electron from the spin-aligned region under the barrier.  
Motivated by the recent momentum-resolved tunneling experiments in a parallel 
wire geometry, we also compute the momentum resolved tunneling matrix elements.  
Our calculations suggest that the eigenstates being expelled are spatially 
localized, consistent with the experimental observations.  However, additional 
mechanisms are needed to account for the experimentally observed large spectral 
weight at near $k=0$ in the tunneling matrix elements.
\end{abstract}
\date{\today}
\maketitle

\section{Introduction}

One-dimensional (1D) electronic systems have proved to be a very 
fruitful field in the studies of interacting many-body systems. The 
infinite homogeneous one-dimensional electron system has been 
extensively studied.  At high density $n\gg a_{B}^{-1}$, where 
$a_{B}=\epsilon\hbar^2/me^2$ is the Bohr radius, the low energy physics 
of the system is well described by the Luttinger model\cite{haldane:81}, 
with spatially extended electronic states as well as separate spin and 
charge excitations propagating at different speeds $v_{s}$ and 
$v_{c}.$\cite{voit:95}  At low density $n\ll a_{B}^{-1}$, a system with 
a long range interaction can be best described as a fluctuating Wigner 
crystal, with electrons being confined around their equilibrium 
positions by their mutual repulsion, though quantum fluctuations prevent 
a true long range order.  The excitations in this case are the density 
fluctuations~(plasmon) of the Wigner crystal and the spinon excitation 
from the Heisenberg antiferromagnetic spin chain created by the exchange 
of the neighboring localized electrons through a barrier formed by their 
mutual repulsions\cite{Matveev:prb}.  For a system with a short range 
interaction, at low densities $n\ll d^{-1}$, where $d$ is the range of 
the interaction, the charge sector can be described as a weakly 
interacting gas of spinless fermions, and the spin sector can again be 
described as a Heisenberg antiferromagnetic chain with an appropriate 
exchange constant $J$.  

Theoretical analyses have also considered the case of a finite wire,
with either sharp or soft confinement at the
ends\cite{yaroslav:02,fiete:05,Mueller:05}.
By contrast, the case of a spatially inhomogeneous 
system, with a low-density region in the center of the wire,  has not been  extensively explored.  
In  recent experiments, Steinberg \textit{et 
al.}\cite{steinberg:06,auslaender:05} used a negatively charged metal 
gate to partially deplete the central region of a finite 
quasi-one-dimensional wire, and studied the low density region by means 
of momentum-conserved tunneling from a parallel ``semi-infinite'' wire 
with higher electron density. They found a striking transition from a 
regime of extended electronic states to a regime of apparently localized 
states, as the electron density at the center of the wire is lowered by 
the negative gate voltage. At a critical value, the electrons  at the 
Fermi level seem to change abruptly from an extended state with 
well-defined momentum into a localized state with a wide range of 
momentum components. In the extended state regime, the tunneling 
measurements show a smooth variation of the electron density in the wire 
as a function of the gate voltage. In contrast, in the localized regime, 
the tunneling only occurs at a series of discrete resonant gate 
voltages, corresponding to the successive expulsion of a single electron 
from a Coulomb blockaded region that is somehow formed under the 
repulsive gate.  Transport measurements  show that the electrical 
conductance along the wire is much smaller than $e^2/h$ when the 
electron-density  under the center gate is low enough to be in the 
localized regime.  Furthermore, measurements of momentum conserved 
tunneling from a second parallel wire show a dramatic change in behavior 
in the localized regime, as we shall discuss further below.

Motivated by the above experiments, we have turned  to the Hartree-Fock 
method to investigate the physical properties of a system of interacting 
electrons on a finite wire with a barrier potential at its center, with 
a special focus on the evolution of the low density electrons.

In previous work, Matveev \cite{Matveev:prb}
studied the case of transport properties of 1D interacting electrons 
through an \emph{adiabatic} barrier, and concluded that the conductance 
is $2e^2/h$ at low temperature and $e^2/h$ at high 
temperature\cite{Matveev:prl}.  However, he did not explore the regime 
where the electron density under the barrier is nearly depleted and the 
two terminal conductance becomes much smaller than $e^2/h$.

Mueller\cite{Mueller:05} explored the crossover from the non-magnetic 
state to the Wigner crystal antiferromagnetic state when reducing the 
electronic density in a finite wire, using a restricted Hartree-Fock 
method. He mostly considered a finite wire that is relatively uniform in 
the center region, under no external magnetic field.  In Appendix B of 
his paper, he briefly considered a wire with an additional potential 
barrier in the center, and found a low density Wigner-crystal like 
regime under the barrier. He did not further investigate the density and 
spin evolution of his system as the density under the barrier is further 
depleted, nor did he study the momentum-dependent tunneling amplitude in 
the case with a low density center region.

Meir and coauthors\cite{meir:prl,meir:nature} studied the formation of 
magnetic moments in a quantum point contact(QPC) in a two-dimensional
geometry using spin-density-functional theory(SDFT). They found that as 
the density inside QPC rises \emph{above} pinch-off, a magnetic moment 
forms inside the opening channel. In longer QPC, the magnetic moments 
take the form of an antiferromagnetically ordered chain. The conducting 
channels inside the QPC can be roughly modelled as a one-dimensional 
system with a smooth potential barrier, and the antiferromagnetic order 
under the barrier they found is consistent with one of the magnetic 
phases we found in our study.  However, unlike the QPC system, our 
strictly one-dimensional system in a strong magnetic field further 
becomes ferromagnetic in low density region near depletion. Furthermore 
in our model we use a modified form of electron interaction to take into 
account of the screening in the tunneling experiments described below.

The paper is organized as following. In Sec.~\ref{model} we introduce 
our basic model, our choice of parameters and the numerical method we 
employed.  In the Sec.~\ref{result}, we present the physical picture of 
successive magnetic phases our system goes through as we increase the 
potential barrier and the crossovers between phases. In Sec.~\ref{wave}, 
we make a more detailed analysis of the form of the wavefunction at 
Fermi energy near depletion. Motivated by the experimental measurements 
by Steinberg \textit{et al.}\cite{steinberg:06,auslaender:05}, we also 
compute the momentum dependence of the tunneling matrix elements for our 
system. In Sec.~\ref{discussion}, we compare our results to the 
Hartree-Fock calculation for an infinite homogeneous system and a 
non-interacting inhomogeneous system, and discuss the implications of 
the electrical conductance measurement from our calculation. In 
Sec.~\ref{summary} we summarize our results and their comparisons with 
experiments.

\section{Model}\label{model}

We consider a system of one-dimensional interacting electrons in a wire 
of length $L=6~\mu m$ with periodic boundary conditions. A uniform 
magnetic  field $B$ is applied throughout the system, which couples only 
to the spins in our model, and which, in most of our calculations, we 
set to $1T$. In the experiments,\cite{steinberg:06,auslaender:05} a 
magnetic field of $1-3T$ was typically applied. 

\begin{figure}
\includegraphics[width=0.5\textwidth]{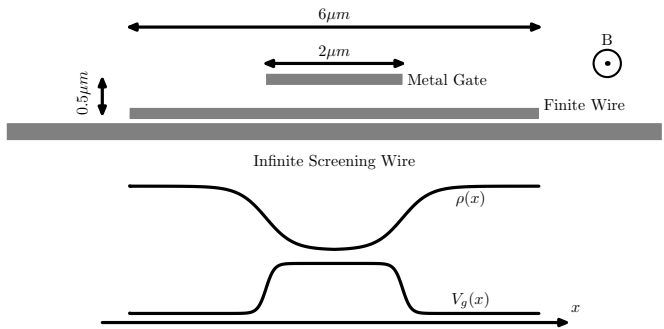}
\caption{\label{fig:exp}A schematic illustration of the geometric 
configurations and the potential and density profiles considered in this 
paper.}
\end{figure}

With the above-mentioned experiments in mind, we assume the electrons 
interact via a Coulomb potential with both a short range and a long 
range cutoff.  The short range cutoff comes from the finite width of the 
experimental wire.  We model it by simply modifying a $1/z$ potential to 
$1/\sqrt{z^2+W^2}$, where $z=x-x'$ is the separation of two electrons 
along the wire and $W$ is the short range cutoff, roughly on the order of half 
of the width of the wire. This density-independent short range cutoff is 
appropriate in our case of a sharp confinement transverse to the direction of 
the wire formed by the cleaved edge overgrowth.   The long range cutoff is the 
result of the screening effect from the higher density wire parallel to the 
short wire in the tunneling experiment.  We model it by putting a second wire, 
which is simplified to be infinitely long and \emph{perfectly conducting}, 
parallel to the finite wire under study, at a center to center distance $d$, as 
shown in Fig.~\ref{fig:exp}.  The resulting form of the interaction $U(z)$ can 
be easily derived, as discussed  in Fiete \textit{et al.}.~\cite{fiete:05}
At short distance $x\ll W$, $U$ levels off smoothly as $1/\sqrt{z^2+W^2}$, 
whereas at long distance $x\gg d$ it decays much more rapidly than Coulomb 
potential $1/z$. Following the experimental setup, in this paper we choose 
$W=0.01~\mu m$ and $d=0.031~\mu m$, and choose the strength of Coulomb 
interaction to correspond the value in the bulk GaAs, yielding a Bohr radius 
$a_B\approx 0.01~\mu m$.

\begin{figure}
\includegraphics[width=0.48\textwidth]{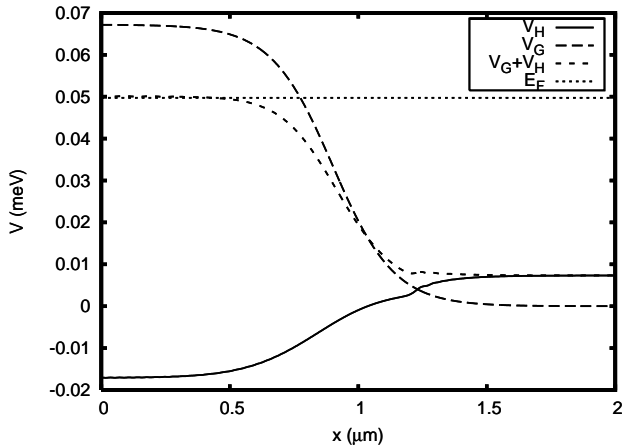}
\caption{\label{fig:potential_barrier}The shape of the bare potential 
barrier $V_{\rm G}$, the Hartree potential $V_{\rm H}$ and their sum, at 
``gate voltage'' $V_{\rm g}=67.2~meV$. The density distribution at this 
gate voltage is shown at the bottom of Fig.~\ref{fig:density_non_anti}.  
All potentials are symmetric around $x=0$, and only the center part 
$x<2~\mu m$ is shown.  The Fermi energy $E_{\rm F}$ of the electrons is 
also shown.}
\end{figure}

In the experiments, a negatively charged 2$~\mu m$ long metal gate at 
0.5 $\mu$m above the finite wire is used to reduce the density at the 
center region of the wire, as illustrated in Fig.~\ref{fig:exp}. To 
approximate the effect of the gate, we use a smooth bare barrier 
potential of the form
\begin{equation}
\label{eq:barrier}
V_{G}(x)=\frac{V_{\rm g}}{1+\exp((|x|-L_{\rm g}/2)/L_{\rm s})}.
\end{equation}
Here $L_{\rm g}$ is the length of the potential barrier, which in our 
calculation we choose to be the length of the experimental metal gate 
$2~\mu m$.  $L_{\rm s}$ controls the sharpness of the edge of the 
potential barrier, which we choose to be on the order of the $0.5~\mu 
m$, the distance from the gate to the finite wire. The quantity $V_{\rm 
g}$ will be referred to, below, as ``gate voltage'', although it is 
actually only proportional to (minus) the applied voltage $V_{\rm G}$.  
The normalization is such that $V_{\rm g}$ is the bare potential at the 
center of the barrier region. The spatial form of this potential can be 
seen in Fig.~\ref{fig:potential_barrier}.

In the experiments, more than one transverse mode in the quantum wires 
can be occupied. Correspondingly, in the wire of our model, we maintain 
two separate subbands of electrons, which we assume to interact only 
through the Hartree term to the electrons in the other subband.  This is 
equivalent to the assumption that in the experimental wire, one can 
ignore any effects of scattering or exchange between electrons in 
different transverse modes. In our model, an energy difference of 
$42meV$ separates the bottom of the two subbands, corresponding to the 
energy separation of the lowest two transverse modes in a square well of 
width $0.02~\mu m$. 

In summary, our Hartree-Fock Hamiltonian can be written as:
\begin{equation}
\label{eqa:Halmintonian}
\begin{split}
H\psi_{\sigma b}(x)=
-\frac{\hbar^{2}}{2m^{*}}
\frac{\partial^2\psi_{\sigma b}(x)}{\partial x^2}
+(V_{\rm{G}}(x)+\Delta_{b}-\frac{1}{2}g^{*}\mu_{B}B\sigma_z)\psi_{\sigma 
b}(x)\\
+V_{H}(x)\psi_{\sigma b}(x)
-\int dx'V_{F}^{\sigma b}(x,x')\psi_{\sigma b}(x')
\end{split}
\end{equation}
\begin{equation}
V_{H}(x)=\int dx'(\sum_{i,\sigma',b'}|\psi_{i\sigma'b'}(x')|^2)
U(x-x')
\end{equation}
\begin{equation}
V_{F}^{\sigma b}=\sum_{i}\psi_{i\sigma b}(x)\psi_{i\sigma b}^{*}
(x')U(x-x').
\end{equation}
Here $m^{*}\approx0.067m_{e}$ is the effective electron mass in bulk $GaAs$, 
$g^{*}\approx0.44$ is the effective $g-$factor in bulk $GaAs$, and $\mu_B$ is 
the Bohr magneton.  $\psi^{*}(x)$ is the complex conjugate of $\psi(x)$.
$\sigma=(\uparrow,\downarrow)$ is spin index.  $b={0,1}$ is the subband 
index: $\Delta_0=0meV$ for $\psi_{\sigma 0}$ in the first subband and 
$\Delta_1=42meV$ for $\psi_{\sigma 1}$ in the second subband. The 
summation over $i$ in computing the Hartree potential $V_H$ and Fock 
kernel $V_F$ is over all the occupied states in a specific spin and 
subband.  Notice, as discussed in the previous paragraph, in computing 
the Fock potential kernel $V_F^{\sigma b}$, we only sum over the 
occupied states with the same spin $\sigma$ and in the same subband $b$ 
as the eigenstate it is acting on.

The numerical method we use in our calculation is a restricted 
Hartree-Fock method~\cite{anderson:92}. 
The electron spins are required to be either parallel or anti-parallel to the applied
magnetic field, so canted spin structures are not allowed.   In the Appendix, 
however, we consider the effects of canting in an infinite homogeneous wire, 
and we argue that canting would have a negligible effect on results for the 
inhomogeneous system, for the parameters of interest to us. 

Starting from solutions to the non-interacting potential barrier 
problem, we iteratively use the Hartree-Fock method until the 
convergence between iterations is achieved. Throughout the calculation, 
we fix the total number of electrons in the finite wire to be $N=1000$, 
whereas the occupation numbers in each spin/subband species remain free 
to change.

In this paper, we focus on the depletion of the first subband under the 
barrier.  In this regime, the second subband is fully depleted under the 
barrier and is only occupied in the outer regions where the total 
density is high.  Consequently, in our calculations, the second subband 
serves  mostly as a reservoir for the electrons under the barrier.  

\section{Magnetic Phases}\label{result}

\begin{figure}
\includegraphics[width=0.48\textwidth]{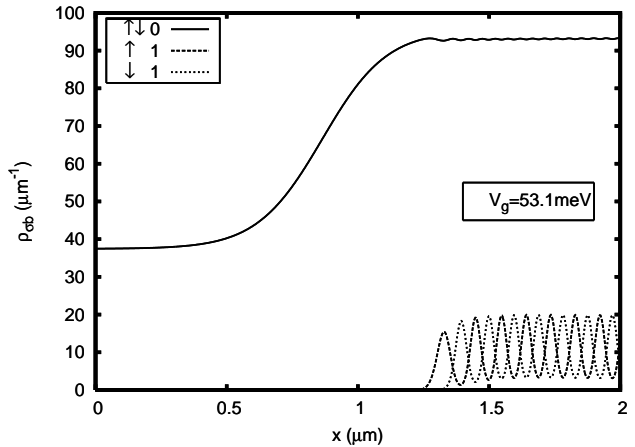}
\caption{\label{fig:density_non}The  density distributions $\rho_{\sigma 
b}(x)$  for the different spin states  and subbands  in the ground 
state, in a high density regime  where the center region is 
non-magnetic.   Identifications of the indices $\sigma, b$ for the 
curves are shown in the inset. For the lower subband ($b=0$), the 
densities of the two spin states are indistinguishable, and are shown by 
a single curve.
All 
densities are symmetric around $x=0$, and only the center part $x<2~\mu 
m$ is shown. }
\end{figure}

\begin{figure}
\includegraphics[width=0.48\textwidth]{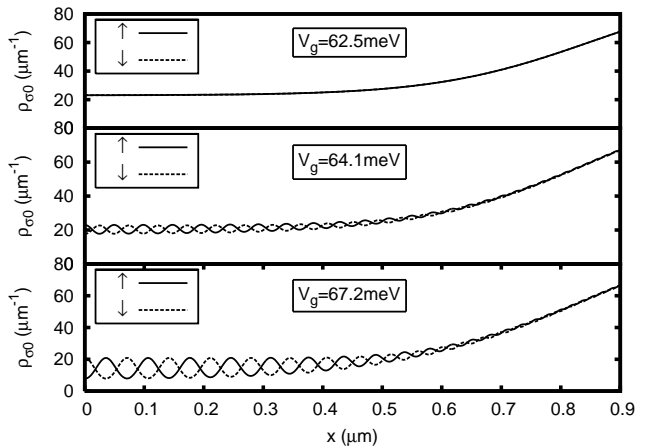}
\caption{\label{fig:density_non_anti} 
Densities of spin-up and spin-down electrons in the center region, showing the emergence of antiferromagnetic order with increasing $V_{\rm{g}}$.
All densities are symmetric around $x=0$, and only the center part of 
the first subband electron densities $x<0.9~\mu m$ is shown.
}
\end{figure}

At high densities, we find that the first subband under the gate is 
essentially unpolarized, as shown in Fig.~\ref{fig:density_non}.
At an electron density $\rho=40\sim50~\mu m^{-1}$, or $\rho_{\sigma 
0}=20\sim25~\mu m^{-1}$ per spin, an
antiferromagnetic order emerges at the low density region under the 
barrier, as shown in Fig.~\ref{fig:density_non_anti}.
The antiferromagnetic order parameter, the staggered magnetization 
$\tilde M$, grows steadily as the density
decreases with the increase of $V_{\rm g}$. There is no sharp transition 
between the nonmagnetic and the antiferromagnetic solutions. 

\begin{figure}
\includegraphics[width=0.48\textwidth]{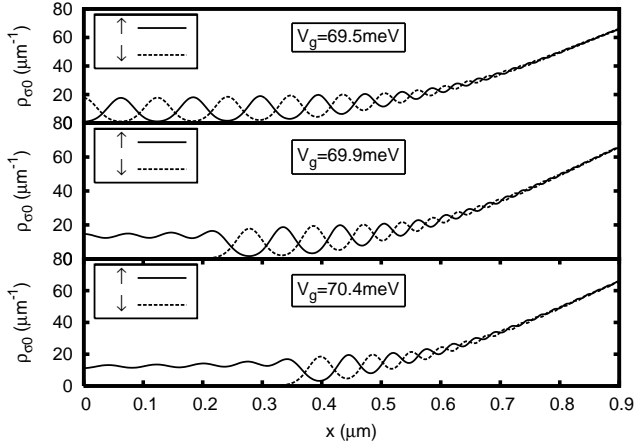}
\caption{\label{fig:density_anti_ferro} The transition from 
antiferromagnetic order to  spin aligned at the center region, in 
the form of a spin-aligned region which expands rapidly  from the center. 
All densities are symmetric around $x=0$, and only the center part of 
the first subband electron densities $x<0.9~\mu m$ is shown.
}
\end{figure}

\begin{figure}
\includegraphics[width=0.48\textwidth]{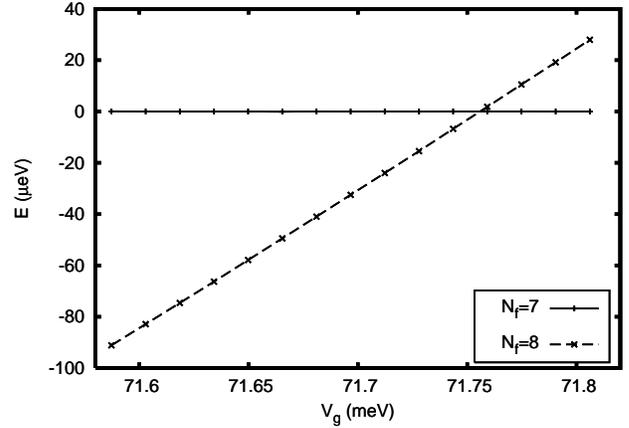}
\caption{\label{fig:transition_energy}This plot shows the crossover, as 
a function of the barrier height $V_{\rm g}$, of the ground state 
energy 
$E$ of two Hartree-Fock solutions, labeled by the number of electrons 
$N_{\rm{f}}$ in spin aligned region at the center.  For the sake of 
clarity, a quadratic function of $V_{\rm g}$ is subtracted from each of 
the ground state. 
}
\end{figure}

\begin{figure}
\includegraphics[width=0.48\textwidth]{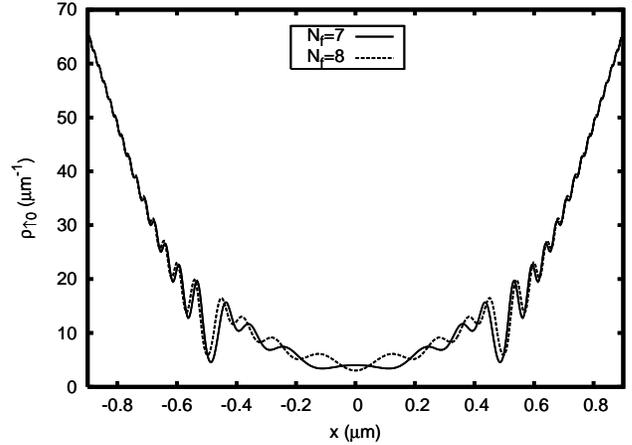}
\caption{\label{fig:transition_density}Electronic density under the 
barrier, before and after the transition shown in 
Fig.~\ref{fig:transition_energy}. Only the densities of first-subband 
spin-up electrons are plotted since electron densities in other states 
do not change greatly during the transition.  The spin density labeled 
of $N_{\rm{f}}=8$ is taken at the $V_{\rm g}=71.74~meV$ right before the 
crossover in Fig.~\ref{fig:transition_energy}, and the one labeled 
$N_{\rm{f}}=7$ is taken at the $V_{\rm g}=71.76~meV$ right after it.}
\end{figure}

\begin{figure}
\includegraphics[width=0.48\textwidth]{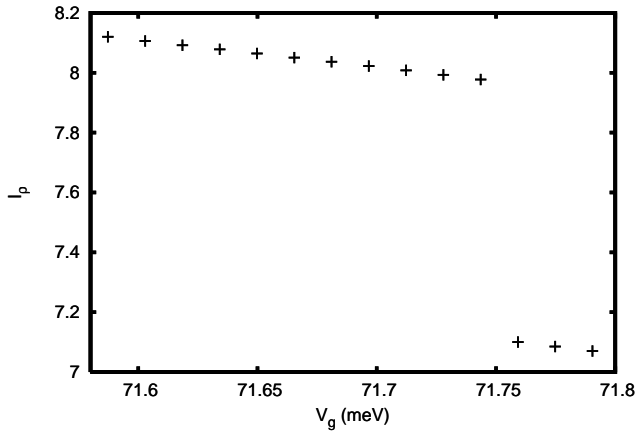}
\caption{\label{fig:expulsion} $I_{\rho}$ is the integrated number of 
spin up electrons in the lower band within the center region defined by 
$-x_{0}\le~x\le~x_{0}$, where $x_{0}=0.498~\mu m$. As a function of the 
gate voltage $V_{\rm g}$, we follow the lowest energy solution across  
the crossover shown in Fig.~\ref{fig:transition_energy}, i.e. the 
$I_{\rho}$ on the left of the crossover is computed from the solution 
$N_{\rm{f}}=8$, and the $I_{\rho}$ on the right from $N_{\rm{f}}=7$.  
The drop in $I_{\rho}$ represents the expulsion of approximately 0.88 
electron from within the center region.
}
\end{figure}

At $\rho=\rho^{*}\sim20~\mu m^{-1}$ under the barrier, a spin aligned 
center region appears and rapidly expands, as shown in 
Fig.~\ref{fig:density_anti_ferro}. As seen in the figure, the spin 
aligned region at the center is sandwiched by two strongly 
antiferromagnetic regions on each side.
Within a narrow range of $V_{\rm g}$, the spin aligned region expands to 
a maximum length, containing $N_{\rm f}=10$ electrons, as shown in the 
bottom of Fig.~\ref{fig:density_anti_ferro}. From that point on to its 
full depletion, the center spin aligned region undergoes a series of 
transitions, each representing the expulsion of one electron from the 
spin aligned region. In contrast with the nonmagnetic and 
antiferromagnetic regimes, where the electronic density $\rho(x)$ under 
the barrier changes smoothly with $V_{\rm g}$, here the $\rho(x)$ in the 
spin aligned center region of the barrier varies discontinuously with an 
increase in $V_{\rm g}$.
Figures~\ref{fig:transition_energy}--~\ref{fig:expulsion} show the 
details of one of such transitions, with the number of electrons in the 
spin aligned region changing from $N_{\rm f}=8$ to $N_{\rm f}=7$ at 
$V_{\rm g}^{*}\approx 71.75~meV$.
In Fig.~\ref{fig:transition_energy}, we see a crossover in the total 
energy of the Hartree-Fock ground state $E$, where the $N_{\rm f}=8$ 
solution has a lower energy for $V_{\rm g}<V_{\rm g}^{*}$ and the 
$N_{\rm f}=7$ solution becomes the ground state for $V_{\rm g}>V_{\rm 
g}^{*}$.  In terms of the occupation numbers of the states of different 
subbands and spins, this transition corresponds to the expulsion of one 
electron from the spin up fist subband to the spin up second subband.  

In general, as $V_{\rm g}$ increases, the transitions in our calculation 
always involve an expulsion of one spin-up first-subband electron to the 
second subband \emph{outside} the center region. But spin flip 
transitions also happen: there are transitions showing the spin-down 
second subband absorbing the expelled electron from spin-up first 
subband.
In Fig.~\ref{fig:transition_density}, the detailed density changes in 
one transition are shown clearly. The spin aligned region in each of the 
solutions is quite well isolated by the antiferromagnetic regions on its 
sides. The first subband spin-down electron density 
$\rho_{\downarrow0}$, which is not shown in 
Fig.~\ref{fig:transition_density} for the sake of clarity, drops steeply 
to zero for $|x|\le 0.45~\mu m$, where $\rho_{\uparrow0}$ rises sharply, 
on both sides of the transition.
The short antiferromagnetic regions on the two sides are only slightly 
shifted in the transition, whereas the center spin-aligned region 
undergoes the significant change from having eight peaks to having 
seven. By plotting the integrated density in the spin-aligned region, 
Fig.~\ref{fig:expulsion} shows that this change in $\rho_{\uparrow0}$ 
indeed amounts to the expulsion of almost a whole electron from the 
region.
The slight deficiency from unity is due to the slightly changed length 
of the spin aligned region, and possibly some small residual density 
from the spin aligned states extending into the antiferromagnetic sides.

It may be possible to detect the existence of the central spin-aligned region 
by the application of a magnetic field $B_{\parallel}$ parallel to the quantum 
wire in question. For a state with $N$ spin aligned electrons in the center 
region, the magnetic field will result a Zeeman energy shift $E_Z=N g^* mu_B 
B_T/2$, where $g^*$ is the $g^*\approx 0.44$ is the effective g-factor in bulk 
$GaAs$, $mu_B$ is the Bohr magneton, and $B_T=\sqrt{B_{\parallel}^2+B^2}$ is 
the strength of the total magnetic field. Due to this energy shift, the 
transition voltage from the $N$ to the $N+1$ spin-aligned electron state will 
be shifted to a larger value, because the $N+1$ electron state energy will be 
lowered by $\Delta E_Z=g^* mu_B/2$ relative to the $N$ electron state. Such a 
shift may be detectable for a large change in the combined field $B$.  For 
example, in the $N=8$ to $N=7$ transition discussed above, a change $\delta 
B_T=4T$ in the combined field will result a shift of $\Delta E_Z\approx 0.092 
meV$.  This is about one fifth of the typical spacing between transition ``gate 
voltage'' $V_{\rm g}^*$, which is approximately $0.45 meV$.

\section{Wavefunctions and Momentum Conserved Tunneling}\label{wave}

\begin{figure}
\includegraphics[width=0.48\textwidth]{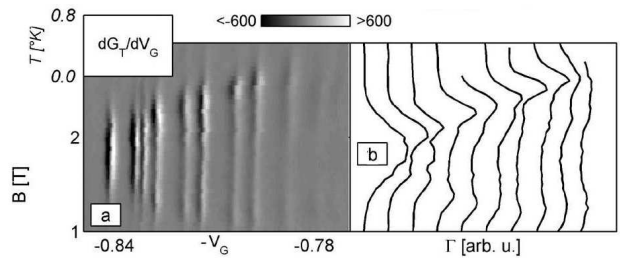}
\caption{\label{fig:matrix_exp}(Color Online) High resolution 
measurement of the 
localized features from Steinberg \textit{et al.}~\cite{steinberg:06}.  
On the left is the raw tunneling data $dG/dV_{\rm{G}}$. On the right are 
the corresponding tunneling rates $\Gamma(B)$, extracted from the 
fitting of the localized features to Coulomb blockade lineshape at each 
magnetic field.  $\Gamma(B)$ is proportional to $|M(k)|^2$, where $M$ is 
defined in Eq.~\ref{equ:matrix} and $B$ controls the momentum of the 
tunneling electron.}
\end{figure}

In the momentum-conserved tunneling experiments of experiments of Steinberg \textit{et al.}, electrons tunnel between 
the finite wire and a parallel ``infinite'' wire, while conserving their
momentum in the wire direction. A magnetic field $B$ perpendicular to
the cleaved edges defining the two quantum wires gives a controllable
momentum boost $q_B=eBd/\hbar$ to the electrons tunneling between the 
wires, where $d$ is the distance between the wires. At low temperature 
and small source-drain bias, the tunneling conductance $G_{T}(B,V_{\rm 
g})\propto~|M|^2$, where the matrix element $M$ has the following physical 
interpretation:\cite{fiete:05,yaroslav:02}
\begin{equation}
\label{equ:matrix}
M=\int_{-\infty}^{\infty}dx~e^{-ikx}~\Psi_{\rm eff}^{N}(x),
\end{equation}
where we have defined a  ``quasi-wavefunction''
\begin{equation}
\label{equ:quasi_wave}
\Psi_{\rm eff}^{N}\equiv \langle N-1| \psi(x)|N\rangle,
\end{equation}
 with  $|N\rangle$ being  the $N$-electron \emph{many-body} ground state 
in the finite wire, $\psi(x)$ being the electron annihilation operator 
at position $x$ in the finite wire, and $k=q_B\pm k_{F}$, where
$k_{F}$ is the Fermi wavevector in the infinite wire. (We neglect, here,  electron-electron interactions in the infinite wire.) The 
momentum dependence of  $|M|$ 
can be extracted from the magnetic-field dependence of the Coulomb blockade 
peak.~\cite{steinberg:06} In the localized regime, expulsion of an electron from the region under the gate is signaled by a vertical stripe in a color plot of the tunnel conductance in the plane of gate-voltage $V_G$ and magnetic field $B$.  (See the left panel of Fig.~\ref{fig:matrix_exp}.)  The momentum dependence of $|M(k)|^2$ is obtained by integrating the  intensity across a given  vertical stripe, at a fixed value of the magnetic field, and comparing the results for different values of  $B$.
As seen in the  
right panel of Fig.~\ref{fig:matrix_exp}, except for the last peak, the 
momentum dependence of the $|M(k)|^2$,  found in the experiments, typically shows to two wide peaks, as well as a
broad momentum distribution between the peaks, signaling relatively localized 
wavefunctions.
\begin{figure}
\includegraphics[width=0.48\textwidth]{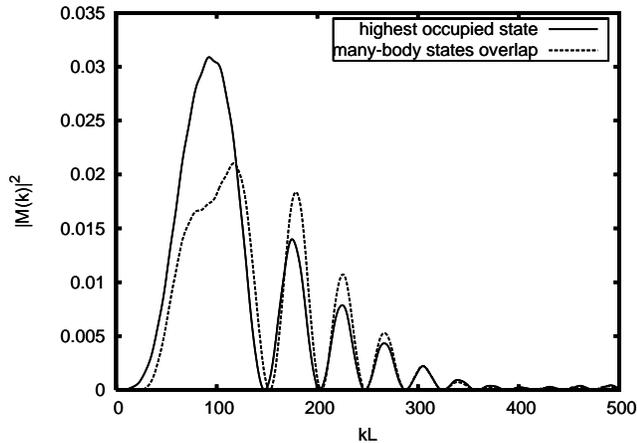}
\caption{\label{fig:matrix_comparison} $|M(k)|^2$ for the transition 
shown in Fig.~\ref{fig:transition_energy}, from $N_{\rm{f}}=8$ to 
$N_{\rm{f}}=7$. The solid line is the squared Fourier transform of the 
wavefunction being expelled. The dotted line is computed from the 
overlap between full Slater determinants.
}
\end{figure}

In our calculation, the $N-1$ and $N$-electron states should be the complete Slater 
determinants of the eigenstates of the corresponding Hartree-Fock 
Hamiltonians. As a simplifying approximation, we may assume  that after expelling one 
electron, the rest of the eigenstates are not affected. In this case 
$\Psi_{\rm eff}^{N}$ is simply the wavefunction of the electron being 
expelled and $M(k)$ is its Fourier transform. $|M(k)|^2$ computed this 
way for the transition $N_{\rm f}=8$ to $N_{\rm f}=7$ is shown as the 
solid line in Fig.~\ref{fig:matrix_comparison}.  The dashed curve is the 
result obtained by using the full Slater determinants of the $(N-1)$ and 
$N$-electron solutions to compute the  matrix elements. This shows an 
orthogonality-catastrophe-type reduction to the overall spectral density 
weight. Both of the calculated  matrix elements show a relatively broad 
momentum distribution, consistent with the experimental result.  
However, the experimental result shows heavy spectral weight near $k=0$,
\emph{between} the two largest peaks at $k=\pm k_{max}$, and little 
weight outside them, whereas our calculations show little weight between 
the largest peaks and considerable weight outside, see 
Fig.~\ref{fig:matrix_comparison}.  This discrepancy suggests that 
additional mechanisms are needed to explain the finer details of 
observed momentum distribution $|M(k)|^2$.

\begin{figure}
\includegraphics[width=0.48\textwidth]{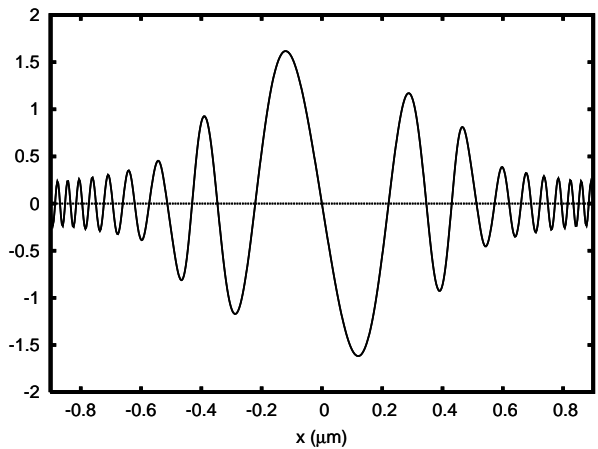}
\caption{\label{fig:wavefunction} This is the highest occupied 
eigenstate of the solution at $V_{\rm g}=71.74~meV$, right before the 
crossover the Fig.~\ref{fig:transition_energy}. The wavefunction is 
normalized and only the center region is plotted.}
\end{figure}

Insight into the calculated shape of  $M(k)$ can be gained by looking at 
the wavefunction in position space for the electron being expelled. 
Fig.~\ref{fig:wavefunction} shows the wave function in position space, 
for which the squared Fourier transform is the solid curve in 
Fig.~\ref{fig:matrix_comparison}. We see that the wave function  has 
relatively large weight in the central spin-aligned region,
roughly for $|x| <0.5~\mu m$.  However, it also has significant weight 
outside the barrier, and in the transition region between.  As will be 
discussed further below, the wave function is qualitatively similar to 
what one would expect in the WKB approximation for a  state slightly 
above the top of a smooth potential barrier.  The separation between 
successive zeroes of the wave function is largest near the center, where 
the amplitude is largest, and it decreases monotonically in the 
transition region, where the amplitude decreases gradually and the 
particle velocity increases. The Fourier transform $M(k)$ has its 
largest amplitude at a value  $|k|=k_m$ corresponding to the spacing 
between zeroes in the center region, and has additional weight at larger 
wavevectors, corresponding to the smaller spacing of zeroes in the 
transition region.  By contrast, the experimental results look like what 
one would find for a particle confined in a soft potential \emph{well}, 
where the zeroes of the wave function would be closest together near the 
center of the well, and be farther apart at the two ends. In this case, 
the Fourier transformed wave functions have significant  weight for 
$|k|$ larger than the peak  value $k_m$, but very little weight at 
larger $|k|$.\cite{yaroslav:02,fiete:05}
In either case, one finds
zeroes in $M(k)$, and oscillations in the amplitude, arising from 
interference between contributions at spatial points $x$ and $-x$.

\section{Discussion}\label{discussion}

\subsection {Hartree-Fock for a homogeneous system}

In order to better understand the results of our calculations, it will be
helpful to recall some of  features of  Hartree-Fock calculations for an
infinite homogeneous one-dimensional electron system.  At high electron
densities, where electron-electron interactions are relatively weak,
with no applied magnetic field,  one finds a nearly-free electron gas,
with a small gap at the Fermi energy, caused by a spin-density wave
corresponding to the wavevector $Q=2k_F$. In position space, this means 
that the unit cell contains precisely two electrons on average, with a 
weak polarization,  alternating between spin up and spin down, along a 
quantization axis that has arbitrary direction\cite{overhauser}.  The 
total charge density will have a period half that of the spin density, 
i.e., there is just one electron in each  charge period.   The amplitude 
of the charge modulation will be proportional to the square of the 
spin-density amplitude, when these modulations are small.  For pure 
Coulomb interactions, the amplitude of the spin-density modulation will 
fall off rapidly at high densities $\rho$, roughly as $\exp (- a\rho)$,  
where $a$ is a length of the order of the Bohr radius. If the 
electron-electron interaction is smooth at short distances,  the 
amplitude of the spin-density modulation can fall off still faster with 
increasing $\rho$.  For a system of  {\em finite} length $L$, we would 
generally not expect to find any spin-density modulation if the 
antiferromagnetic coherence length, which is inversely proportional
to spin-density amplitude of the infinite system, becomes larger than $L$.

As the electron density is lowered the amplitudes of the spin and
charge modulations both grow until one reaches  the situation of a
Wigner crystal, where the there is strong modulation in the charge
density, and there is nearly complete spin polarization, alternating
up and down for successive electrons.  At still lower densities,
the Hartree-Fock approximation predicts a phase transition to a fully
aligned ferromagnetic state.  The ferromagnetic state is, of course,
an artifact of the Hartree-Fock approximation, as it is known\cite{lieb} 
that the exact ground state is a spin singlet, for $B=0$. However, a 
state of full spin alignment should occur at low densities, for $B \neq 
0$, and the Hartree-Fock approximation may be a reasonable description 
of this transition for $B=$1T.

The predicted antiferromagnetic order is also an artifact of the
Hartree-Fock approximation, as quantum fluctuations would be expected to
replace  the long-range spin order with correlations that fall off as a
power of the distance. In an infinite system, charge density modulations
will also be destroyed by quantum fluctuations of the positions of the
electrons on the Wigner crystal.  However, these quantum fluctuations can
be relatively weak when there is strong repulsion between the electrons,
and a significant charge density modulation may exist in a finite system
of moderate length.

Hartree-Fock calculations for an infinite system, with the same
interaction potential used in our finite system, are presented  in the
Appendix.  We discuss there also the effect of spin-canting in an 
applied magnetic field.

According to Fig.~\ref {fig:unrestricted_energy} of the Appendix, the
Hartree-Fock transition to a fully spin-polarized state should occur at a
density of approximately 16 electrons per micron for an infinite system
with the parameters of the model under consideration, in a Zeeman field
$B=$1T. This is similar to the density $\rho^* \approx 20$nm$^{-1}$,
where we found the onset of a center region with  full spin alignment,
in our calculations for the system with a barrier.

\subsection{Inhomogeneous system without interactions}

It is also useful to review what one would expect for an inhomogeneous
system analogous to the wire under consideration, but without
electron-electron interactions. In particular, we may consider what should
happen as one varies the height of a smooth center barrier, similar
to the bare potential barrier in Fig.~\ref {fig:potential_barrier} or
to the self-consistent potential in that figure, including the Hartree
potential but not the non-local exchange potential.  By comparing this
qualitative picture with the  results of our Hartree-Fock calculations,
we can better see whether there are features of the latter which reflect
in an essential way the exchange and correlation features of a strongly
interacting many-electron system.

It is important to note that the total length of our system is finite,
so there will be a discrete set of energy levels for the system as
a whole.  As the  total length is six microns, and the flat potential
area under the gate is of order one to two microns, the majority of the
length is outside the barrier region.  Because the density of states
in a non-interacting one-dimensional system is inversely proportional
to the electron density $\rho$, however, the local density of states for
electrons in the barrier region can be much higher than the density of
states outside, if the electron density is sufficiently low in the 
barrier region. Thus, if the chemical potential is fixed, and the height 
of the barrier is lowered below the Fermi energy, we may expect to see a 
closely spaced sequence of electrons entering into states whose 
amplitudes are highly concentrated in the center region.  Indeed, if the 
overall system length is not too large, we may expect a large fraction 
of the probability density for each added electron will be located in 
the barrier region.

It should be noted that the upper wire in the experiments  of
Ref.~[\onlinecite{steinberg:06}] has an overall  length that is not
very different from the wire used in our calculations.  The experimental
wire is not truly isolated, but is tunnel-coupled to leads in its outer
regions; so its energy levels should actually be life-time broadened. If
the escape rate from the wire is smaller than the spacing between energy
levels, however, the features of a finite system should be maintained.

For non-interacting electrons, if the total number of particles is fixed,
rather than the chemical potential, then the electrons entering the center
must be transferred from electron states outside the barrier region,
(e.g., states belonging to a second  subband), and the Fermi level will
itself decrease each time an electron is added to the center region.
If the level spacing of the outside bands is larger than the level spacing
of the center region, then the spacing between  gate potentials where
successive electrons enter the center region will be determined by the
larger energy spacing between these reservoir states.

We may also consider what would happen if one had an {\em infinite}
system of non-interacting electrons, with a flat barrier of finite
length in the center.  Suppose the potential $V(x)$ is zero outside the
barrier region and equal to $V_{\rm g}$ at the center of the barrier. If 
the barrier is smooth enough so that one can use the WKB approximation,
then the wavefunction $\psi(x)$ for a state with energy $E$ slightly
above $V_{\rm g}$ will have an amplitude which is larger inside the 
central region than outside, by a factor $[(E - V(x)) / E]^{-1/4} 
\approx
[\rho_0 / \rho(x)]^{1/2}$, where $\rho(x)$ is the cumulative 
(Thomas-Fermi)
electron density at point $x$ from all states with energy less than $E$,
and $\rho_0$ is the electron density far from the barrier.   For a 
smooth
barrier, we see that $\rho(x)$, and hence the amplitude of the 
wavefunction
$\psi$, should have a maximum in the center of the barrier, and fall
off monotonically with increasing $|x|$. The spacing between successive
zeroes of $\psi$ should be given by $1/\rho(x)$, which should decrease
monotonically with increasing $|x|$.  We note that the Hartree-Fock
wavefunction plotted  in Fig.~\ref{fig:wavefunction} is qualitatively
consistent with these features.

If the barrier height is varied continuously at fixed Fermi energy, for an
infinite system, when the WKB approximation is valid, the wavefunctions
will vary continuously, and the number of particles above the barrier
will likewise vary in a continuous fashion The WKB approximation will
break down, however,  if the energy $E$ gets too close to the top of
the barrier.  For a smooth potential, one expects the WKB approximation
to be valid for all but the last one or two states above the barrier.
By contrast, for a flat-topped potential that falls off relatively
abruptly at the ends of the barrier, deviations may more pronounced.
In this case we may find a number of resonant states above the barrier,
which have very small electron density outside the barrier region and
which exist only in narrow energy bands.  Then, as the gate voltage is
varied, the number of electrons in the barrier region will increase by one
in a narrow region of gate voltage, as each resonance passes through  the
Fermi energy.  In the limit of a potential  $V(r)$  that drops sharply
to $V = - \infty$ at the edge of the barrier, a wavefunction $\psi(x)$
with weight inside  the barrier region will vanish outside this region;
the wavefunctions and energy levels will be discrete and will be the same
as if there was an infinitely high potential at the end of the barrier.

\begin{figure}
\includegraphics[width=0.48\textwidth]{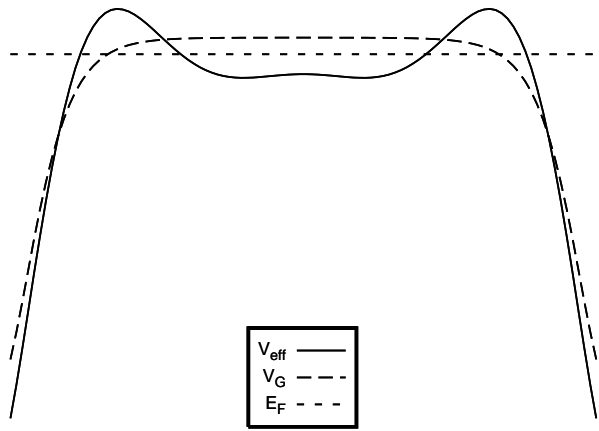}
\caption{\label{fig:schematic}A schematic plot of an effective barrier 
potential $V_{\rm eff}$ that could produce a form of the wavefunction 
more consistent with the results of the tunneling experiments. The 
dashed line is the bare potential barrier $V_{\rm G}$ and the dotted 
line is the Fermi level $E_{\rm F}$.}
\end{figure}

We may also imagine  a situation where the self-consistent potential
$V_{\rm eff}(x)$ is smooth but non-monotonic, having somehow developed a 
pair of
maxima near the edges of the original barrier region, as illustrated in
Fig~\ref{fig:schematic}. In this case, there will   be an energy range 
such that $E$ is smaller than the maximum value of $V_{\rm eff}$ but 
larger than the value at the center of the barrier region.  If this 
energy range is large enough,
there may be a discrete series of states which are well localized in the
classically allowed region, decay to a small value in the classically
forbidden regions, and have only a small amplitude outside the barrier.
The distance between zeroes of the wavefunction will then increase
with increasing $|x|$ in  the region where the wavefunction is large.
The form  of $|M(k)|^2$ that one would obtain by taking the Fourier
transform of this wavefunction will have a maximum intensity at a wave
vector corresponding to the local Fermi wave vector $k_F$ at $x=0$, and
will have significant weight for  $|k| < k_F$, but very little weight at
$|k| > k_F$. This result is qualitatively similar to the observations  of
Steinberg et al., illustrated in Fig.~\ref  {fig:matrix_exp}. However,
it is quite different from what we have obtained from our Hartree Fock
calculations, illustrated in Fig.~\ref{fig:matrix_comparison}, where
there is considerable weight at large $k$.

The spin-density structures obtained in our  Hartree-Fock calculations
suggest that a non-monotonic self-consistent potential, similar to
that in Fig.~\ref{fig:schematic}, might in fact have been a 
qualitatively reasonable
representation of the Hamiltonian seen by the electrons with the
majority spin orientation. Because there is a fully polarized spin-down
electron on either side of the central region of spin-up electrons ,
and because there is a strong repulsion between electrons of  opposite
spin, one might expect that the spin up electrons would see an effective
potential maximum at the positions of the spin-down electrons.  The strong
antiferromagnetic order just outside this region, suggests that there
might be, locally, an antiferromagnetic energy gap at the Fermi energy,
so that the wavefunction of  the highest energy filled state would decay
further as it passes through this region.  Our numerical results suggest
however, that these effects, if they are present, are not strong enough
to produce the type of localization one might have hoped for.

The qualitative resemblance between the calculated Hartree-Fock
wavefunction and the WKB form  for states above a smooth potential
barrier, discussed above, appears to persist down to densities
where there are only one or two electrons left at the top of the
barrier.

\subsection{Electrical conductance}

A striking feature of the experimental results is the occurrence of a
sharp drop in the conductance through the finite wire in a regime where 
there
were still several electrons in the region below the center gate.(Here 
we refer to the two terminal conductance $G$ measured through contacts 
at two ends of the finite wire, and not the tunnelling conductance 
$G_{\rm T}$ for current flowing between the finite wire and the 
semi-infinite wire). For a
one-dimensional system of non-interacting electrons, in a potential $V(x)$
that vanishes outside a central region, one can relate the electrical
conductance, using the Landauer-Buttiker formula, to the transmission
probability for an incident electron at the Fermi energy.  For a potential
barrier which is symmetric under reflection, the transmission probability,
in turn, can be expressed in terms of the phase shifts  for states of
even and odd parity.  In the Hartree-Fock approximation, however,  this
analysis is complicated by several factors.  Although electron-electron
interactions in the leads are relatively weak because of the high electron
density there, they would still give rise to antiferromagnetic order,
in a lead of infinite length, as discussed above.  Thus,  in principle
there should always be an energy gap at the Fermi energy, and phase shifts
cannot be defined.  In practice, this should not be a serious problem for
our system, because the calculated energy gap is extremely small at high
densities, and one could estimate the conductance from phase shifts at
energies outside of the energy gap, but still close to the Fermi energy.

A more significant problem arises from the fact that our computations
use a system of finite length, and we have only a discrete set of energy levels.
Analyzing these wave functions, we may obtain even-parity and odd-parity
phase shifts at a discrete set of energies, but we do not obtain both even
and odd phase shifts at a single energy.  We can obtain some estimate
of the phase shifts for an infinite system, however,  by looking at
the alternation between energy levels for even and odd numbered wave
functions of a given spin and band index, in the finite system.

For an energy high above the barrier, we expect that  WKB is a good
approximation, which means that an incident particle is transmitted with
essentially no reflection, corresponding to a conductance of $e^2/h$
per spin.  This means that there is no difference in the phase shifts
for even and odd parity. In our model calculations, we assume periodic
boundary conditions at the ends of the wire. Then, for a large but finite
system, when there is negligible reflection at the barrier, we expect
energy levels to occur in pairs, with even and odd parity states that
are nearly degenerate.

For an energy well below the barrier, where there is nearly total
reflection, the even and odd parity phase shifts should differ by
approximately $\pi/2$.  Then, with periodic boundary conditions, we
expect energy levels to alternate between even and odd parity states,
with nearly equal spacings between them.

The energy spacings we find in our Hartree-Fock calculations are in good
agreement with these expectations provided the Fermi level is not too
close to the barrier top. Thus we have near perfect transmission when
the Fermi level is well above the barrier, and near perfect reflection
when it is well below.   However, we have not been able to analyze
the conductivity in the most interesting region, when there are only a
few electrons in the spin-polarized region at the top of the barrier,
essentially because our system size is too small, and we do not have
enough energy levels in that region.

Finally, we note that the Landauer-Buttiker conductance discussed above
applies to a wire that is connected to its leads by adiabatic, 
non-reflecting contacts. In the experiments by Steinberg \textit{et 
al.}\cite{steinberg:06,auslaender:05}, the contacts from the finite wire 
to the two dimensional electron gas(2DEG) are not perfectly adiabatic, 
and will add contact resistance to any resistance discussed above.

\section{Summary}\label{summary}

In summary, from our Hartree-Fock calculations, we have developed
a picture of successive magnetic phases in the low density region
of an inhomogeneous one-dimensional electron systems in a uniform
magnetic field. The depleted electrons under the barrier first enter an
antiferromagnetic phase, then, near depletion, part of the lowest density
electrons become spin aligned and get isolated from the high density
region outside the barrier by two antiferromagnetic regions sandwiching
it. The final stage of depletion takes the form of successive expulsion
of a single electron from the spin aligned region, resulting successive
periods of relative insensitivity of the spin aligned electron density
to $V_{\rm g}$, followed by the sudden rearrangement due to the 
expulsion
of one electron.

\begin{figure}
\includegraphics[width=0.48\textwidth]{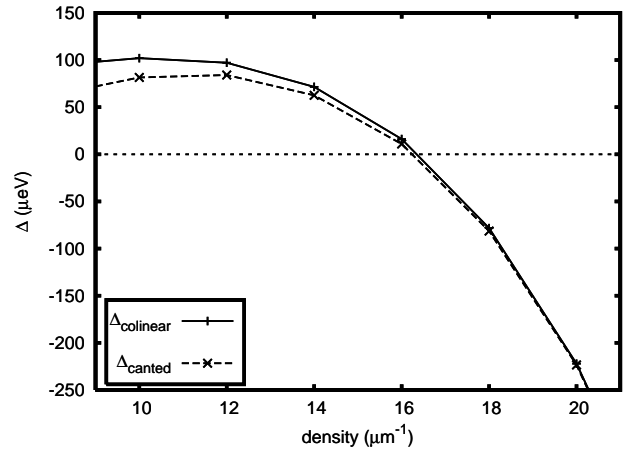}
\caption{\label{fig:unrestricted_energy} The energy difference per 
electron $\Delta$ between the antiferromagnetic and the fully spin 
aligned solutions in an infinite uniform electron system, as a function 
of electron density. Both systems are in a uniform magnetic field 
$B=1T$.  The dotted curve is the canted solution and the solid curve is 
the solution with spin colinear to $B$.}
\end{figure}

The most serious discrepancy between our calculations and the experimental
results of Steinberg et al is the form factor for momentum-conserved
tunneling in the regime where there are of the order of four to  ten
electrons under the central gate.  Our matrix elements have too much
weight at large momenta, It is not clear what is the source of this
discrepancy, However, it may be that potential  fluctuations due to
residual disorder are important.  A small random potential due to
charged impurities set back from the wire may have little effect on
the mean free path for relatively high electron densities, but could 
lead to strong back scattering and localization at very low densities, 
where the kinetic energy can be small, and where the small value of 
$2k_F$ permits backscattering from potential fluctuations of relatively 
long wave length.

As mentioned earlier, calculations near  the depletion of the upper band
at a smaller $V_{\rm g}$,  give results similar to those obtained
near the depletion of the lowest band. We find, again, a spin-aligned
central region, sandwiched by antiferromagnetic regions on each sides.
A similar phenomenon of sudden expulsions of a single localized electron
from the spin aligned parts are also observed. This is consistent with
the experimental observation of similar localization behavior upon
the depletion of the second subband~\cite{steinberg:06}.  Small potential
fluctuations due to impurities may again be important for explaining
the experimental results for momentum-dependent tunneling in this regime.

\subsection*{Acknowledgments}
We would like to thank Hadar Steiberg, Ophir Auslaender, Amir Yacoby, 
Yaroslav Tserkovnyak and Greg Fiete for illuminating discussions. We
have also  benefitted from discussions with Walter Hofstetter and Gergely 
Zarand when working on a related earlier project. This work is supported by NSF 
grants DMR05-41988 and PHY06-46094.

\section*{APPENDIX SPIN-CANTING IN THE HOMOGENEOUS SYSTEM}
\begin{figure}
\includegraphics[width=0.48\textwidth]{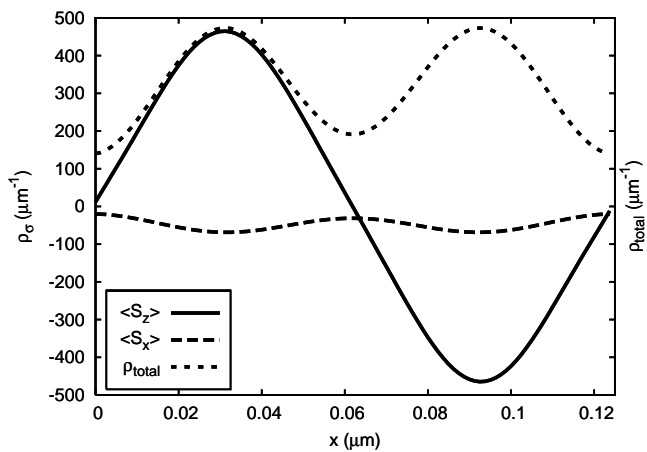}
\caption{\label{fig:unrestricted_density}The spin polarization in $x$ 
and $z$ direction in the canted solution at density $\rho=16~\mu 
m^{-1}$, right at the transition from the antiferromagnetic to the 
spin-aligned ground state. Due to the magnetic field in $x$ direction, a 
small paramagnetic component is developed in addition to the dominant 
antiferromagnetic magnetization in $z$ direction. Only a single unit 
cell, consisting of two electrons, are shown. The total density is also 
shown as the dotted line.}
\end{figure}

Our calculations for the system with a barrier have been carried out using  a restricted Hartree-Fock 
method, where  canted spin states were not allowed. For a classical 
Heisenberg model in an applied uniform field $\vec{B}$,  the  antiferromagnetic state will spontaneously align  itself so that the staggered spin component is perpendicular to the applied field, and the  individual spins will cant towards the 
direction of the applied field. Hence,  it may be asked whether 
allowing canting of the spins  in a Hartree-Fock calculation would
significantly change our results. To clarify this issue we have  carried out 
both restricted and unrestricted Hartree-Fock calculations on a uniform 
\emph{infinite} electron system in a magnetic field. The interaction 
between electrons is the same as we used before, and the magnetic field 
is also $B=1T$. As shown in Fig.~\ref{fig:unrestricted_energy} and 
Fig.~\ref{fig:unrestricted_density}, the effects of canting in terms of 
both the energy gains and the changes in spin density are small.
Canting is only relevant to our zero-temperature calculation when the 
antiferromagnetic solution is the ground states. Even near the 
transition density $\rho^{*}=16~\mu m^{-1}$ from the antiferromagnetic 
ground state to the fully spin aligned ground state, when canting is the 
greatest, the energy gained by allowing canting is only $2.64~\mu eV$ 
per electron.  

Extrapolating these results to the system with a barrier, we see that even if we allowed ten antiferromagnetic electrons on each 
side of the solution shown in Fig~\ref{fig:transition_density} to cant, 
the total energy gain would be minuscule compared with the level spacing at 
Fermi energy there.  At $\rho^{*}$, in the homogeneous system,  the magnetization parallel to the uniform 
magnetic field, in the canted state,  is less than one tenth of the 
antiferromagnetic magnetization perpendicular to it. Thus, our calculations 
for  the homogeneous system suggest that allowing canting in the Hartree-Fock 
calculation of the inhomogeneous one dimensional wire would not give 
qualitatively different results from our restricted calculations above.

We remark that for a homogeneous system in zero-magnetic field, with the
same electron-electron interaction as above, the onset of ferromagnetism
in the Hartree-Fock approximation would occur at a density $\rho^*
\approx 14~\mu m^{-1}$.


\bigskip



\begin{thebibliography}{48}
\expandafter\ifx\csname natexlab\endcsname\relax\def\natexlab#1{#1}\fi
\expandafter\ifx\csname bibnamefont\endcsname\relax
  \def\bibnamefont#1{#1}\fi
\expandafter\ifx\csname bibfnamefont\endcsname\relax
  \def\bibfnamefont#1{#1}\fi
\expandafter\ifx\csname citenamefont\endcsname\relax
  \def\citenamefont#1{#1}\fi
\expandafter\ifx\csname url\endcsname\relax
  \def\url#1{\texttt{#1}}\fi
\expandafter\ifx\csname urlprefix\endcsname\relax\def\urlprefix{URL }\fi
\providecommand{\bibinfo}[2]{#2}
\providecommand{\eprint}[2][]{\url{#2}}

\bibitem{haldane:81}
\bibinfo{author}{\bibfnamefont{F.~D.~M.} \bibnamefont{Haldane}},
  \bibinfo{journal}{J. Phys. C.} \textbf{\bibinfo{volume}{14}},
  \bibinfo{pages}{2585} (\bibinfo{year}{1981}).

\bibitem{voit:95}
\bibinfo{author}{\bibfnamefont{J.} \bibnamefont{Voit}},
  \bibinfo{journal}{Rep. Prog. Phys.} \textbf{\bibinfo{volume}{58}},
  \bibinfo{pages}{977} (\bibinfo{year}{1995}).

\bibitem{Matveev:prb}
\bibinfo{author}{\bibfnamefont{K.~A.} \bibnamefont{Matveev}},
  \bibinfo{journal}{Phys. Rev. B} \textbf{\bibinfo{volume}{70}},
  \bibinfo{pages}{245319} (\bibinfo{year}{2004}).

\bibitem{yaroslav:02}
\bibinfo{author}{\bibfnamefont{Y.} \bibnamefont{Tserkovnyak}}, 
\bibinfo{author}{\bibfnamefont{B.~I.} \bibnamefont{Halperin}}, 
\bibinfo{author}{\bibfnamefont{O.~M.} \bibnamefont{Auslaender}}, 
\bibnamefont{and}
\bibinfo{author}{\bibfnamefont{A.} \bibnamefont{Yacoby}}, 
\bibinfo{journal}{Phys. Rev. Lett.} \textbf{\bibinfo{volume}{89}},
  \bibinfo{pages}{136805} (\bibinfo{year}{2002}).

\bibitem{fiete:05}
\bibinfo{author}{\bibfnamefont{Gregory A.} \bibnamefont{Fiete}},
\bibinfo{author}{\bibfnamefont{Jiang} \bibnamefont{Qian}},
\bibinfo{author}{\bibfnamefont{Yaroslav} \bibnamefont{Tserkovnyak}},
\bibnamefont{and}
\bibinfo{author}{\bibfnamefont{Bertrand.~I} \bibnamefont{Halperin}},
\bibinfo{journal}{Phys. Rev. B.} \textbf{\bibinfo{volume}{72}},
\bibinfo{pages}{045315} (\bibinfo{year}{2005}).

\bibitem{Mueller:05}
\bibinfo{author}{\bibfnamefont{E.~J.} \bibnamefont{Mueller}},
  \bibinfo{journal}{Phys. Rev. B} \textbf{\bibinfo{volume}{72}},
  \bibinfo{pages}{75322} (\bibinfo{year}{2005}).

\bibitem{auslaender:05}
\bibinfo{author}{\bibfnamefont{O.~M.} \bibnamefont{Auslaender}}, 
    \bibinfo{author}{\bibfnamefont{O.~M.} \bibnamefont{Steinberg}},
    \bibinfo{author}{\bibfnamefont{A.} \bibnamefont{Yacoby}}, 
\bibinfo{author}{\bibfnamefont{Y.} \bibnamefont{Tserkovnyak}}, 
\bibinfo{author}{\bibfnamefont{B.~I.} \bibnamefont{Halperin}}, 
\bibinfo{author}{\bibfnamefont{K.~W.} \bibnamefont{Baldwin}}, 
\bibinfo{author}{\bibfnamefont{L.~N.} \bibnamefont{Pfeiffer}}, 
\bibnamefont{and}
  \bibinfo{author}{\bibfnamefont{K.~W}~\bibnamefont{West}},
  \bibinfo{journal}{Science} \textbf{\bibinfo{volume}{308}},
  \bibinfo{pages}{88} (\bibinfo{year}{2005}).

\bibitem{steinberg:06}
\bibinfo{author}{\bibfnamefont{H.} \bibnamefont{Steinberg}},
    \bibinfo{author}{\bibfnamefont{O.~M.} \bibnamefont{Auslaender}}, 
    \bibinfo{author}{\bibfnamefont{A.} \bibnamefont{Yacoby}}, 
    \bibinfo{author}{\bibfnamefont{J.} \bibnamefont{Qian}}, 
    \bibinfo{author}{\bibfnamefont{G.~A.} \bibnamefont{Fiete}}, 
    \bibinfo{author}{\bibfnamefont{Y.} \bibnamefont{Tserkovnyak}}, 
    \bibinfo{author}{\bibfnamefont{B.~I.} \bibnamefont{Halperin}}, 
    \bibinfo{author}{\bibfnamefont{K.~W.} \bibnamefont{Baldwin}}, 
    \bibinfo{author}{\bibfnamefont{L.~N.} \bibnamefont{Pfeiffer}}, 
\bibnamefont{and}
  \bibinfo{author}{\bibfnamefont{K.~W}~\bibnamefont{West}},
  \bibinfo{journal}{Phys. Rev. B.} \textbf{\bibinfo{volume}{73}},
  \bibinfo{pages}{113307} (\bibinfo{year}{2006}).

\bibitem{Matveev:prl}
\bibinfo{author}{\bibfnamefont{K.~A.} \bibnamefont{Matveev}},
  \bibinfo{journal}{Phys. Rev. Lett.} \textbf{\bibinfo{volume}{92}},
  \bibinfo{pages}{106801} (\bibinfo{year}{2004}).

\bibitem{meir:prl}
\bibinfo{author}{\bibfnamefont{Y.} \bibnamefont{Meir}},
    \bibinfo{author}{\bibfnamefont{K.} \bibnamefont{Hirose}},
    \bibnamefont{and}
    \bibinfo{author}{\bibfnamefont{N.S.} \bibnamefont{Wingreen}},
  \bibinfo{journal}{Phys. Rev. Lett.} \textbf{\bibinfo{volume}{89}},
  \bibinfo{pages}{196802} (\bibinfo{year}{2002}).

\bibitem{meir:nature}
\bibinfo{author}{\bibfnamefont{T.} \bibnamefont{Rejec}},
    \bibnamefont{and}
    \bibinfo{author}{\bibfnamefont{Y.} \bibnamefont{Meir}},
  \bibinfo{journal}{Nature} \textbf{\bibinfo{volume}{442}},
  \bibinfo{pages}{900} (\bibinfo{year}{2006}).


\bibitem{anderson:92}
\bibinfo{author}{\bibfnamefont{P.~W.} \bibnamefont{Anderson}},
\emph{\bibinfo{title}{Concepts in Solids}} 
(\bibinfo{publisher}{Addison-Wesley Publishing Co. Inc.}, 
\bibinfo{year}{1992}).

\bibitem{overhauser}
\bibinfo{author}{\bibfnamefont{A. W.} \bibnamefont{Overhauser}},
\bibinfo{journal}{Phys. Rev.} \textbf{\bibinfo{volume}{128}},
\bibinfo{pages}{1437} (\bibinfo{year}{1962}).

\bibitem{lieb}
\bibinfo{author}{\bibfnamefont{E.} \bibnamefont{Lieb}},
\bibnamefont{and}
\bibinfo{author}{\bibfnamefont{D} \bibnamefont{Mattis}},
\bibinfo{journal}{Phys. Rev.} \textbf{\bibinfo{volume}{125}},
\bibinfo{pages}{164} (\bibinfo{year}{1962}).

\end{thebibliography}
\end{document}